# Robust Adaptive Control for Circadian Dynamics: Poincaré Approach to Backstepping Method

**Myroslav Sparavalo**
NYC Transit Authority, Telecommunications Division
126 53rd Street, NY, NY 10019, USA
mksparavalo@yahoo.com

## ABSTRACT
A mathematical model of the circadian dynamics in the form of Van der Pol equation with an external force as a control is investigated. The combination of backstepping method and differential-topological techniques based on the Poincaré's ideas is used. The robust model identification adaptive control for a specific adaptation law is designed.

## INTRODUCTION
This paper has the objective to implement differential-topological tools of the study of complex nonlinear dynamic systems for the analysis and synthesis of human-machine systems. The core of the model describing the jet lag and the process of adaptation after rapid trans-meridian travel is a Van der Pol equation with an external force. Consider it written in polar coordinates as follows

$$\frac{dr}{dt} = \mu r \cos^2 \theta \left(1 - r^2 \sin^2 \theta\right) + u \cos \theta$$
$$\frac{d\theta}{dt} = 1 - \mu \sin \theta \cos \theta \left(1 - r^2 \sin^2 \theta\right) - \frac{\sin \theta}{r} u, \quad (1)$$

where $r$ and $\theta$ are phase variables; $t$ is time; $\mu \in M \subset \mathbb{R}$ is a system scalar parameter; $u \in [u_{\min}; u_{\max}]$ is an scalar external force or control and $u = u(r,\theta) \in \mathbb{C}^1(\mathbb{R}^2 \to \mathbb{R}^1)$.

The control aim is to find a terminal feedback control law $u = \hat{u}(r,\theta,a)$ bringing the system (1) into a terminal manifold as we will call the periodic-in-polar-coordinates curve of the form

$$M = \{r = g(\theta,b)\}, \quad (2)$$

where $g(\theta,b) = g(\theta + 2\pi n, b), n \in Z$, $a = (a_1, a_2) \in A \subset \mathbb{R}^2$ and $b = (b_1, b_2) \in B \subset \mathbb{R}^2$ are the vectors of parameters of control and terminal manifold respectively. Let them belong to the space of piecewise constant functions. Designate $a^0 = (a_1^0, a_2^0) \in A, b^0 = (b_1^0, b_2^0) \in B, \mu^0 \in M$ some nominal values of $a, b, \mu$. There are three additional requirements to the sought control $u = \hat{u}(r,\theta,a)$. It should be stable in the large, robust and adaptive according to certain adaptation law, which will be defined later, taking into consideration the unpredictable changes of the parameters of the terminal manifold $b$ and the system $\mu$.

At control designing we will use the combination of the backstepping method [1] and the Poincaré approach to the investigation of ordinary differential equations with controlling functions in their right-hand sides [2].

### STEP 1 – CHANGING PHASE VARIABLES

$$r = v + g(\theta,b), \theta = \theta; \frac{dr}{dt} = \frac{dv}{dt} + \frac{\partial g(\theta,b)}{\partial \theta}\frac{d\theta}{dt}, \frac{d\theta}{dt} = \frac{d\theta}{dt}.$$

We receive a new transformed system

$$\frac{dv}{dt} = P(v,\theta,u,\mu,b)$$
$$\frac{d\theta}{dt} = 1 - \mu \sin \theta \cos \theta \left(1 - (v + g(\theta))^2 \sin^2 \theta\right) - \frac{\sin \theta}{(v + g(\theta))} u,$$

where

$$P(v,\theta,u,\mu,b) = u\left(\cos \theta + \frac{\partial g(\theta,b)}{\partial \theta}\frac{\sin \theta}{(v + g(\theta,b))}\right) +$$
$$+ \mu \cos^2 \theta (v + g(\theta,b))\left[1 - (v + g(\theta,b))^2 \sin^2 \theta\right] +$$
$$+ \mu \sin \theta \cos \theta \frac{\partial g(\theta,b)}{\partial \theta}\left[1 - (v + g(\theta,b))^2 \sin^2 \theta\right] - \frac{\partial g(\theta,b)}{\partial \theta}.$$

### STEP 2 - FORMING THE RIGHT-HAND SIDE OF THE DIFFERENTIAL EQUATION
We set the equation
$$P(v,\theta,u,\mu,b) = \chi(v,\theta,a).$$

Solving it for $u$, we obtain the sought control law in the form
$$u = u'(v,\theta,a,b,\mu),$$
where

$$u'(v,\theta,a,b,\mu) = \frac{\chi(v,\theta,a) - \Gamma_1' \cdot \mu \cos \theta \cdot \Gamma_2' + \frac{\partial g(\theta,b)}{\partial \theta}}{\cos \theta + \frac{\partial g(\theta,b)}{\partial \theta}\frac{\sin \theta}{(v + g(\theta,b))}}$$

with $\Gamma_1'$ and $\Gamma_2'$ as

$$\Gamma_1' = 1 - (v + g(\theta,b))^2 \sin^2 \theta,$$
$$\Gamma_2' = \cos \theta (v + g(\theta,b)) + \sin \theta \frac{\partial g(\theta,b)}{\partial \theta}.$$

### STEP 3 - RETURNING TO OLD PHASE VARIABLES
Using the inverse transformation $\{v = r - g(\theta,b), \theta = \theta\}$ we

return to the initial system (3) with the sought control law as follows

$$u = \bar{u}(r,\theta,a,b,\mu) = \left\{\cos\theta + \frac{\partial g(\theta,b)}{\partial \theta}\frac{\sin\theta}{r}\right\}^{-1} \times$$
$$\times \left\{\chi(r - g(\theta,b),\theta,a) - \Gamma_1 \cdot \mu\cos\theta \cdot \Gamma_2 + \frac{\partial g(\theta,b)}{\partial \theta}\right\}, \quad (3)$$

where $\Gamma_1 = 1 - r^2 \sin^2\theta, \Gamma_2 = r\cdot\cos\theta + \sin\theta\frac{\partial g(\theta)}{\partial \theta}$ and at that $\bar{u}(r,\theta,a,b,\mu) =|r = v + g(\theta)| = u'(v,\theta,a,b,\mu)$.

## STEP 4 - CHOOSING THE FUNCTION $\chi(v,\theta,a)$

According to [3] to ensure robustness and stability in the large the function $\chi(v,\theta,a)$ should be associated with the covering mapping $\pi_v$ belonging to the class of the covering maps $[\pi_v]^A$ with $rank\pi_v = 1$, which generate structurally-stable attractors in phase spaces. The parametric function

$$\chi = \chi(v,\theta,a) = -a_1 \cdot \cos^2\theta \cdot \arctan(a_2 v) \quad (4)$$

satisfies the above-mentioned condition and allows to take account of the constraints imposed on the control $u$ to boot.

## STEP 5 - COMPUTER SIMULATION IN MATLAB

Let the system parameter be $\mu = 0.1$, the terminal manifold (2) be $M = \{r = g(\theta) = b_0 + b_1\sin(\theta)\}$ with the parameters $b_0 = 4, b_1 = 1$ and control parameters be $a_1 = 0.5, a_2 = 1$. The computer simulation of the system (1) with the control law (3), the terminal manifold $M$ and the parameters defined above in MATLAB illustrated with Figures 1 and 2.

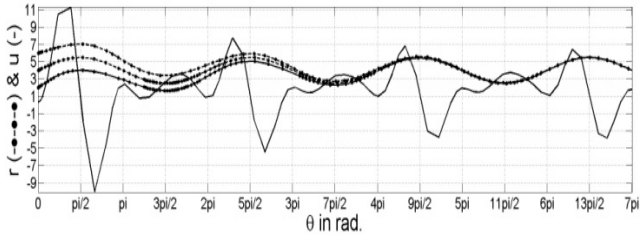

FIGURE 1. THE PHASE PORTRAIT OF THE FORCED VAD DER POL SYSTEM IN POLAR COORDINATES WITH THE TERMINAL MANIFOLD r = 4 + 1.5·sin(θ) AND THE GRAPH OF CONTROL u AS A FUNCTION OF θ

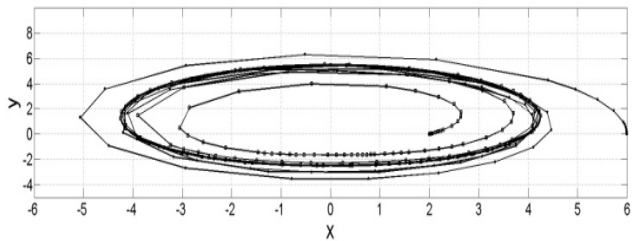

FIGURE 2. THE PHASE PORTRAIT OF THE FORCED VAN DER POL SYSTEM IN CARTESIAN COORDINATES WITH THE TERMINAL MANIFOLD (x² + y² - 1.5 y)² - 16 (x² + y²) = 0 IDENTICAL TO r = 4 + 1.5·sin(θ) IN POLAR ONES

The surface representing the function (4) with the specific parameters $a_1 = 0.5, a_2 = 1$ is given below.

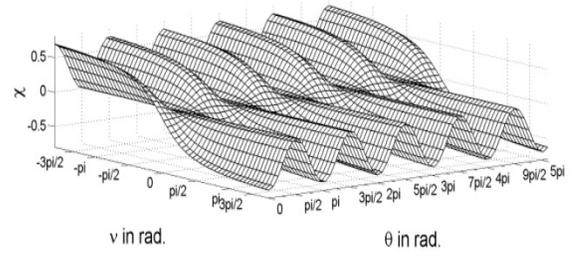

FIGURE 3. THE GRAPH OF THE SURFACE $\chi = -0.5 \cos(\theta)^2 \text{ atan}(v)$

## STEP 6 - MAKING THE CONTROL ADAPTIVE

We will understand the adaptation in the sense of the ability of the control to adapt to the unpredictable changes of the parameters of the terminal manifold, the plant and external perturbations by satisfying some adaptation law in the form of an equality/inequality or maintaining the extreme of some functional related to control quality. So we deal with Model Identification Adaptive Control (MIAC) [3].

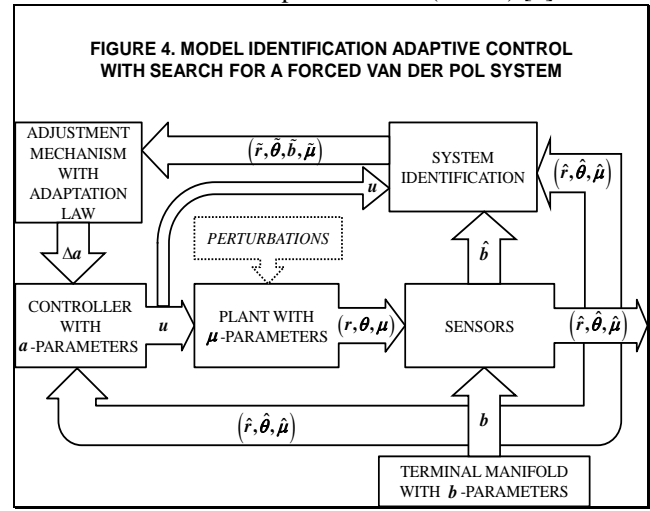

FIGURE 4. MODEL IDENTIFICATION ADAPTIVE CONTROL WITH SEARCH FOR A FORCED VAN DER POL SYSTEM

Here $(\hat{r},\hat{\theta},\hat{b},\hat{\mu})$ and $(\tilde{r},\tilde{\theta},\tilde{\mu},\tilde{b})$ are the estimates and the identified values of $(r,\theta,\mu,b)$ output by the sensors and the system identification respectively; $\Delta a$ is some correction to $a^0$ produced by the adjustment mechanism. Let the adaptation law be the requirement to satisfy the constraint on the control $u \in [u_{\min}; u_{\max}]$ at any admissible variations of the parameters $b = (b_1,b_2) \in B, \mu \in M$. According to (3) this adaptation law can be expressed as

$$\exists a = a^0 + \Delta a \in A : u_{\min} \leq \bar{u}(\tilde{r},\tilde{\theta},a,\tilde{b},\tilde{\mu}) \leq u_{\max}.$$

## REFERENCES
1. Kristić, M., Kanellakopoulos, I., Kokotović, P. Nonlinear and Adaptive Control Design. John Wiley & Sons, Inc., 1995.
2. Sparavalo, M. A Method of Goal-Oriented Formation of Local Topological Structure of Co-Dimension One Foliations for Dynamic Systems with Control, J. of Automation & Information Sci., 25(2), 1992, 65-71.
3. B. H. C. Cheng et al. (Eds.): Software Engineering for Self-Adaptive Systems, LNCS 5525, pp.48-70, 2009.